\documentclass[12pt]{article}
\usepackage{graphicx}
\usepackage{pazh}
\usepackage{lscape}
\usepackage{amssymb}
\usepackage[T2A]{fontenc}  
\usepackage[koi8-r]{inputenc}
\usepackage[english]{babel}

\tightenlines

\def\arcsec{$^{\prime\prime\,}$}
\def\arcmin{$^{\prime\,}$}

\def\а{$^{\mbox{\small а}}$}
\def\б{$^{\mbox{\small б}}$}
\def\в{$^{\mbox{\small в}}$}
\def\г{$^{\mbox{\small г}}$}

\begin{document}

\title{\bf  XTE J1901+014 the First Low-Mass Fast X-ray Transient? }
\author{\bf \hspace{-1.3cm}\copyright\,2008 г. \ \
D.I. Karasev\affilmark{1}$^{\,*}$,
A.A.Lutovinov\affilmark{1},
R.A.Burenin\affilmark{1}}
\affil{
$^1$ {\it Space Research Institute, ul. Profsoyuznaya 84/32, Moscow, 117997 Russia} \\
}

\vspace{2mm}

\centerline{Received June 17, 2008}

\sloppy
\vspace{2mm}
\noindent

\sloppy


             We continue to study the fast X-ray transient XTE J1901+014 discovered in 2002 by the
RXTE observatory, whose nature has not yet been established. Based on the XMM-Newton observations
of the source in 2006, we have obtained its energy spectrum, light curves, and power spectrum in the
energy range 0.5-12 keV, which are in good agreement with our results obtained previously from the
data of other observatories. In turn, this suggests that the source's emission is stable in the quiescent
state. The XMM-Newton observations also allowed the source's localization accuracy to be improved
to <2'' , which subsequently enabled us to search for its optical companion with the RTT-150 and 6-m
BTA (Special Astrophysical Observatory) telescopes. Combining optical, X-ray, and infrared observations,
we have concluded that the optical companion in the system under study can be either a later-type star at a
distance of several kpc or a very distant red giant or an A or F star. Thus, XTE J1901+014 may be the first
low-mass fast X-ray transient.

{\bf Key words:\/} fast X-ray transients, outbursts, neutron stars, black holes.

\vfill

{$^{*}$ e-mail: dkarasev@hea.iki.rssi.ru}

\vfill
{PACS: 98.70.Qy}

\section*{INTRODUCTION}

    XTE J1901+014 was discovered by the all-sky monitor (ASM) onboard the RXTE space observatory during its outburst on April 6, 2002, with a peak flux of 1.2 Crab in the energy range 1.5-12 keV that lasted from $> 2$ min to $< 3.15$ h. The coordinates of the source (J2000.0), R.A. = 19$^h$ 01$^m$ 46$^s$ and DEC = +1\deg 24\arcmin 15\arcsec.7 ., were determined with an accuracy of $\approx$3\arcmin(Remillard and Smith 2002).  
                         
Only one persistent X-ray source, 1RXS J190141.0+012618, from the ROSAT catalog falls within the error box in this direction. Based on INTEGRAL, RXTE, and ROSAT data, Karasev et al. (2007) concluded that (1) these sources are identical and (2) a persistent flux of $\approx$ 2.7 mCrab in the energy range 0.6-100 keV is recorded from XTE J1901+014 in the quiescent state. Karasev et al. (2007) searched for the possible optical companions of the source and assumed that this is a low-mass system, which is atypical of fast X-ray transients. Analysis of the archival ASM data provides evidence for the latter. It revealed a previous outburst
from the object under study in July 1997 with a peak
flux of 0.4-0.5 Crab in the energy range 1.5-12 keV
and a duration > 6 min and < 8 hours (Remillard and
Smith 2002).

Note that none of the outbursts could be associated with the X-ray afterglow of gamma-ray bursts, since there were no bursts in the corresponding catalogs close in time and position to the outbursts under study. No other outbursts of such an intensity have been detected from the source. 

Based on XMM-Newton data, Smith et al. (2007) improved the localization accuracy of XTE J1901+014 to $\sim$1\arcsec. Subsequent infrared observations of this sky region with the 3.5-m TNG telescope (La Palma) did not reveal any optical companion in the J and H bands. A very weak signal was detected only in the K band (Smith et al. 2007).

     In this paper, we reanalyzed the XMM-Newton observations of XTE J1901+014 in October 2006 and searched for the optical companion with the RTT-150 and BTA (Special Astrophysical Observatory) telescopes. Combining our results with the data from the optical USNO-B1 and GSC 2.3 and infrared 2MASS and DENIS catalogs, we concluded that the optical companion in the system is highly likely to be a low mass star.

\section*{OBSERVATIONS}
We used the data from the PN and MOS X-ray telescopes onboard the XMM-Newton observatory obtained when XTE J1901+014 was observed on October 14, 2006. The source's effective exposure time was $\approx$ 5 ks. The data were processed with the standard SAS 7.1.0 software (http://xmm2.esac.esa.int/sas/). For the subsequent analysis of the X-ray (0.5-12 keV) spectra and light curves of the source, we used the software included in the standard HEASOFT package (http://heasarc.gsfc.nasa.gov/docs/software/lheasoft/).

To search for the optical companion, we used data from well-known optical and infrared catalogs, USNO-B1 and 2MASS, as well as additional observations of the source with the Russian-Turkish RTT-150 telescope and the BTA telescope (Special Astrophysical Observatory). Upper limits for the possible optical companion were estimated with RTT-150 in the r'-band (SDSS) and with BTA in the I-band. The data from the optical telescopes were processed with the software included in the IRAF package (http://iraf.noao.edu/) using the corresponding photometric solutions, The total RTT-150 and BTA exposure times were $\approx$3600 and $\approx$600 s, respectively.

\section*{SPECTRAL AND TIMING ANALYSES}
Based on XMM-Newton data, we were able to obtain a high-quality, statistically significant spectrum of XTE J1901+014 in the energy range 0.5-12 keV, which was analyzed using the XSPEC package. The source's spectrum is well fitted by the simplest power-law model with absorption,
 
$$ A(E)=K\times (E/1keV)^{-\Gamma}$$

where $\Gamma$ is the photon index and K is the normalization of photon keV$^{-1}$cm$^{-2}$s$^{-1}$ to 1 keV. The absorption is specified by the function $M(E) = exp(-N_{H}\times \sigma(E)))$, where $\sigma(E)$ is the absorption cross section (Morrison and McCammon 1983), and the atomic hydrogen column density toward the source is $N_{H}=(2.58\pm0.11)\times 10^{22}$ atom cm$^{-2}$ ; the corresponding photon index and 0.6-12 keV flux are  $\Gamma= 1.98\pm0.03$ and $\approx$ 2.5 mCrab, respectively (Fig. 1). These results are in good agreement with the results of the spectral analysis obtained by Karasev et al.(2007) using a combination of data from the other observatories in a wide energy range (0.5-100 keV). This leads us to conclude that the source's emission in the quiescent state is spectrally persistent. Note that the results of our spectral analysis based on MOS and PN data agree closely.

It is well known that the spectra of low-mass neutron-star binaries can be fitted by a combination of two blackbody spectra, one of which represents the accretion-disk emission and the other represents the boundary-layer emission (see, e.g., Sibgatullin and Sunyaev 2000). Applying this model to the description of the spectrum for the object under study yields the following best-fit parameters: $N_{H}=1.4\times 10^{22}$ atom cm$^{-2}$, $kT1 = 0.82\pm0.03$ keV, and $kT2 = 2.57\pm0.05$ keV. These are in good agreement with the observed boundary-layer and accretion-disk temperatures (see, e.g., Gilfanov and Revnivtsev 2005).

To reveal temporal features, we constructed the source's power spectrum using PN/XMM-Newton data. The power spectrum revealed no periodic variability of the flux from the source (Fig. 2). However, according to Titarchuk et al. (2007), a slope of $\approx$1.5 is typical of accreting systems. In addition, the power spectrum exhibits an implicit feature (flattening) near a frequency of 0.001 Hz, which, according to the results of the same paper, can be associated with a hot corona.                                             

The XMM-Newton data are also indicative of an aperiodic variability in the system, which spectrally corresponds to thequiescent state, except for the normalization. This was pointed out previously by Karasev et al. (2007) based on RXTE data.

\section*{LOCALIZATION AND THE OPTICAL COMPANION}
XTE J1901+014 was at the center of the PN and MOS felds of view during the XMM-Newton observations. The statistical error in the source's position, $\approx$ 0\arcsec.3, was determined using a standard procedure of the searching for and localizing sources. The typical XMM-Newton systematic error in the source position is $\approx$ 2\arcsec (1$\sigma$). Thus, based on XMM-Newton data, we localized XTE J1901+014 with the coordinates R.A. = 19$^h$ 01$^m$ 40$^s$.22 and Dec = +01\deg 26\arcmin25\arcsec.37 (J2000) and an accuracy of $\approx$2\arcsec. Note that these coordinates coincide, within the error limits, with those of Smith et al. (2007), but the localization accuracy for the latter is better ($\approx$1\arcsec). The improved position of the source slightly differs from its position found by Karasev et al. (2007) using HRI/ROSAT data. Previously, this caused diffculties with the identification of the optical companion (see Fig. 3a).

Subsequent observations of this region with the optical RTT-150 and BTA (Special Astrophysical Observatory) telescopes failed to unambiguously determine the optical companion of the source due to its faintness (see Figs. 3a and 3b). We managed to obtain only upper limits for the magnitudes in the corresponding optical bands. The upper limits in the r' (RTT-150) and I (BTA) bands are $\sim$23.5 and $\sim$24.5 mag, respectively. The derived limits and the absence of an infrared signal (see, e.g., Fig. 3c) suggest that the system under consideration is unlikely to be a high-mass X-ray binary in our Galaxy, which is consistent with the assumption made by Smith et al. (2007).

In our case, the uncertainty in the interstellar absorption toward the source severely hampers a more accurate determination of the possible type of the system. Therefore, we made absorption estimates using stars close to XTE J1901+014, more specifically, those lying within 4\arcmin of XTE J1901+014.

Since the colors of stars depend only on the absorption, we determined the types of the neighboring stars based on the color tables from Bonneau et al. (2006) by varying the absorption and using 2MASS, DENIS, USNO-B1, and GSC 2.3 data.
Subsequently, using the K(J-K) color diagram from Zurita Heras et al. (2008), we chose the distance in such a way that the K magnitude typical of a star of a given type corresponded to the value of J-K obtained above. Figure 4 presents the distance dependence of the absorption obtained by the method described above. Here, the absorption error is determined by the step with which we vary the absorption, while the distance error is related to the errors on the K(J-K) diagram. The figure presents the results for $\sim$30 stars. Since the magnitudes from the USNO-B1 and GSC catalogs have fairly large errors ($\sim$0.15-0.3 mag), an uncertainty in determining the type of stars arises, i.e.,depending on the absorption, several types of stars correspond to the same set of magnitudes from the catalog and, accordingly, we obtain several points in the figure for the same combination of magnitudes from the catalog. In addition, the type of the star may not change with absorption. In this case, the crosses in Fig. 4 form vertical trends. It follows from the
 constructed distance dependence of the absorption that $N_H$ does not exceed $0.8\times10^{22}$ atom cm$^{-2}$ for the most distant stars (4-5 kpc). These results are in satisfactory agreement with the data obtained from the H I absorption maps (Dickey and Lockman 1990) in this direction, $(0.6-0.8)\times10^{22}$ atom cm$^{-2}$.

 Having estimated the interstellar absorption toward the source and the upper limits in two optical bands, let us now attempt to estimate the possible types of the stars as the optical companion and the distance to the system. It is pointed out in the telegram by Smith et al. (2007) that a small excess of the signal above noise was detected in the XTE J1901+014 error box with the 3.5-m TNG telescope in the infrared K band, while no excess was observed in the J and H bands. Therefore, for the subsequent analysis, we assume that the upper limit for the source's K magnitude is 14.5, which was obtained from the 2MASS and DENIS data.

    In the subsequent analysis, by assuming the stellar spectrum to be a blackbody one, we constructed the dependence of the star's distance on its surface temperature so as to satisfy the derived upper limits on the colors. Because of the well-known uncertainty in the type-luminosity relation, we took the maximum and minimum luminosities for each type of stars to estimate the upper and lower limits on the distance to the system. The analysis was performed for the cases with different absorptions, $N_H = 0.8\times10^{22}$ atom cm$^{-2}$ (maximum interstellar absorption in this direction), $N_H = 1.4\times10^{22}$ atom cm$^{-2}$ (this value was obtained by fitting the X-ray spectrum of the source by a combination of blackbodies), and $N_H = 2.58\times10^{22}$ atom cm$^{-2}$ (a power-law fit to the spectrum). The results are presented on Fig. 5. Here, the long dashed line marks the distance to the Galactic edge in this direction ($\sim$17-20 kpc). Thus,we see from Fig. 5 that a star from the class of OB giants, typical of fast X-ray transients, cannot be the companion of the system for any absorption consistent with the X-ray data. A star of late G-K types is a possible optical companion of the system. In this case, additional absorption can arise either at the edge of an inclined accretion disk or from inhomogeneities of the interstellar medium on the line of sight. Another possible companion is a very distant A or F star. In this case, the additional absorption can exist in the system due to stellar wind, but the source becomes a super-Eddington one during its outbursts. Figure 5 (short dashed line) shows the distance at which the source during its 2002 outburst would reach the Eddington luminosity limit for the neutron star. In our case, this is $\sim$5 kpc.
\section*{CONCLUSIONS}                           
In this paper, based on X-ray, optical, and infrared data, we carried out studies to establish the nature of the source XTE J1901+014.

The source's power spectrum is typical of accreting X-ray binaries with a slope of $\approx$-1.5 and reveals no periodic variability of the flux.
                                                          
Based on XMM-Newton data, we obtained improved coordinates of the source. Subsequently, this allowed us to perform observations of this sky region with the RTT-150 and BTA (Special Astrophysical Observatory) telescopes to find the optical companion. The observations yielded only uppers limits on the r' and I magnitudes (23.5 and 24.5 mag, respectively). In combination with the absence of a statistically significant infrared signal (Smith et al. 2007), this suggests that the system should be a low-mass one.
                                                          
A serious difficulty for the ultimate identification of the system and the determination of its distance is a significant excess of the absorption derived from the X-ray spectrum (($1.4-2.6)\times10^{22}$ atom cm$^{-2}$ , depending on the spectral model used) compared to the interstellar absorption toward the source $(0.6-0.8)\times 10^{22}$ atom cm$^{-2}$. Note that despite the excess absorption in the spectrum, the source cannot be an absorbed high-mass system, i.e., a system located in a cloud of gas and dust, since, in this case, the absorptions derived from the X-ray spectra turn out to be considerably larger (see, e.g., Lutovinov et al.2005a). Furthermore, as was pointed out by Smith et al. (2007), an infrared excess should have been detected from the region under study in this case, but it is not observed.

     For example, a red giant as the optical companion, absorption in the corona above the edges of an inclined accretion disk, etc. can be the possible explanations of the enhanced absorption. The so-called seed photons coming from the neutron star surface, having a certain minimum energy, and then Comptonized in the hot corona can also be responsible for the observed excess absorption. However, the absorption associated with this effect is smaller than the observed excess of absorption in the spectrum of XTE J1901+014. Finally, the enhanced absorption may be associated with local inhomogeneities toward the source, for example, it may stem from the fact that a small molecular or gas-dust cloud falls on the line of sight by chance.
  
The distance to the source also remains uncertain,since it strongly depends on the system type (see Fig. 5). If the source's luminosity during its outbursts is assumed to reach the Eddington limit, then the distance to the object is $\sim$5 kpc. If the system is located at this distance or closer, then only a late-type star can be the optical companion (Fig. 5); if, alternatively, the system is much farther, then a late A or F main-sequence star can act as the opticalcompanion. In this case, however, the source during its outbursts is essentially a super-Eddington one, much as is observed for V4641 Sgr (Revnivtsev et al.2002). The latter is also possible if the companion in the system is a distant red giant (Fig. 5c).

In conclusion, note that XTE J1901+014 differs significantly from other known fast X-ray transients, such as IGR J17544-2619 (Sunyaev et al. 2003), XTE J1739-302 (Smith et al. 2006; Lutovinov et al.2005b), AX J1749.1-2733 (Karasev et al. 2008), SAXJ1818.6-1703 (Grebenev and Sunyaev 2005) etc.; it is probably the first fast X-ray transient in a low-mass X-ray binary. Note also the interesting similarity between XTE J1901+014 and Swift J195509.6+261406, which, just as XTE J1901+014, was also initially classified as a gamma-ray burst (Kasliwal et al. 2008).

\section*{ACKNOWLEDGMENTS}
    We are grateful to V.A. Arefiev, S.A. Grebenev, M.G. Revnivtsev, and E.V. Filippova for discussions of the results. This work was supported by the Russian Foundation for Basic Research (project nos. 07-02-01051 and 07-02-01004), the Presidium of the Russian Academy of Sciences (the ``Origin and Evolution of Stars and Galaxies'' Program), the OFN-
17 Program, and the Presidential Program for Support of Leading Scientific Schools (project no. NSh-5579.2008.2).

\section*{REFERENCES}
1. D. Bonneau, J. M. Clausse, X. Delfosse, et al., Astron. Astrophys. 456, 789 (2006).

2. J. M. Dickey and F. J. Lockman, Astron. Astrophys. 28, 215 (1990).

3. M. Gilfanov and M. Revnivtsev, Astron. Nachrichten 326, 812 (2005).

4. S. A. Grebenev and R. A. Sunyaev, Astron. Lett. 31, 672 (2005).

5. M. Kasliwal, S. Cenko, S. Kulkarni S., et al., Astrophys. J. 678, 1127 (2008).
6. D. I. Karasev, A. A. Lutovinov, and S. A. Grebenev, Pis'ma Astron. Zh. 33, 186 (2007) [Astron. Lett. 33, 159 (2007).

7. D. Karasev, S. Tsygankov, and A. Lutovinov, Mon.Not. R. Astron. Soc. 386, 10 (2008).

8. D. J. Kirkpatrick, N. I. Reid, J. Liebert, et al., Astrophys. J. 519, 802 (1999).

9. A. Lutovinov, M. Revnivtsev, M. Gilfanov, et al., Astron. Astrophys. 444, 821 (2005a).  
                     
10. A. Lutovinov, M. Revnivtsev, S. Molkov, and R. Sunyaev, Astron. Astrophys. 430, 997 (2005b).

11. R. Remillard and D. Smith, Astron. Telegram. 88, 1 (2002).

12. M. Revnivtsev, M. Gilfanov, E. Churazov, and R. Sunyaev, Astron. Astrophys. 391, 1013 (2002).

13. M. Revnivtsev, S. Sazonov, M. Gilfanov, et al., Astron. Astrophys. 452, 169 (2006).

14. N. R. Sibgatullin and R. A. Sunyaev, Astron. Lett. 26, 699 (2000).

15. S. Smith, W. Heindl, C. Markwardt, et al., Astrophys.J. 638, 974 (2007).

16. D. Smith, R. Rampy, and I. Negueruela, Astron. Telegram. 1268, 1 (2007).

17. R. Sunyaev, S. Grebenev, A. Lutovinov, et al., Astron.Telegram. 190, 1 (2003).

18. L. Titarchuk, N. Shaposhnikov, and V. Arefiev, Astrophys. J. 660, 334 (2007).

19. J. A. Zurita Heras and S. Chaty, Astron. Astrophys.,in press (2008).

\newpage
\begin{figure*}
\includegraphics[width=17cm]{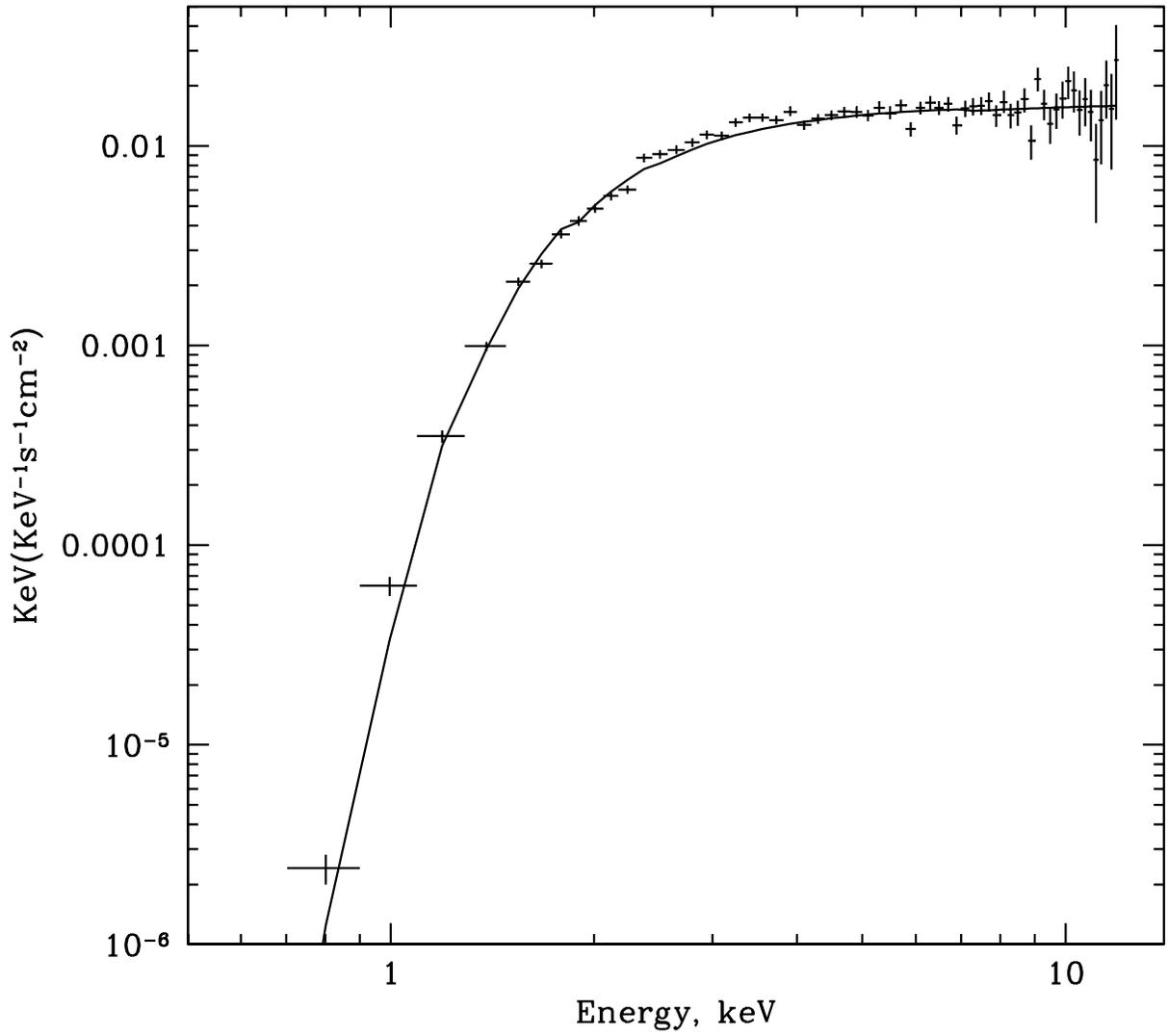}
\caption{Energy spectrum of XTEJ1901+014 in the energy range 0.5-10 keV obtained from PN/XMM-Newton data and fitted by a power-law with interstellar absorption.}

\end{figure*}

\newpage

\begin{figure*}
\includegraphics[width=17cm]{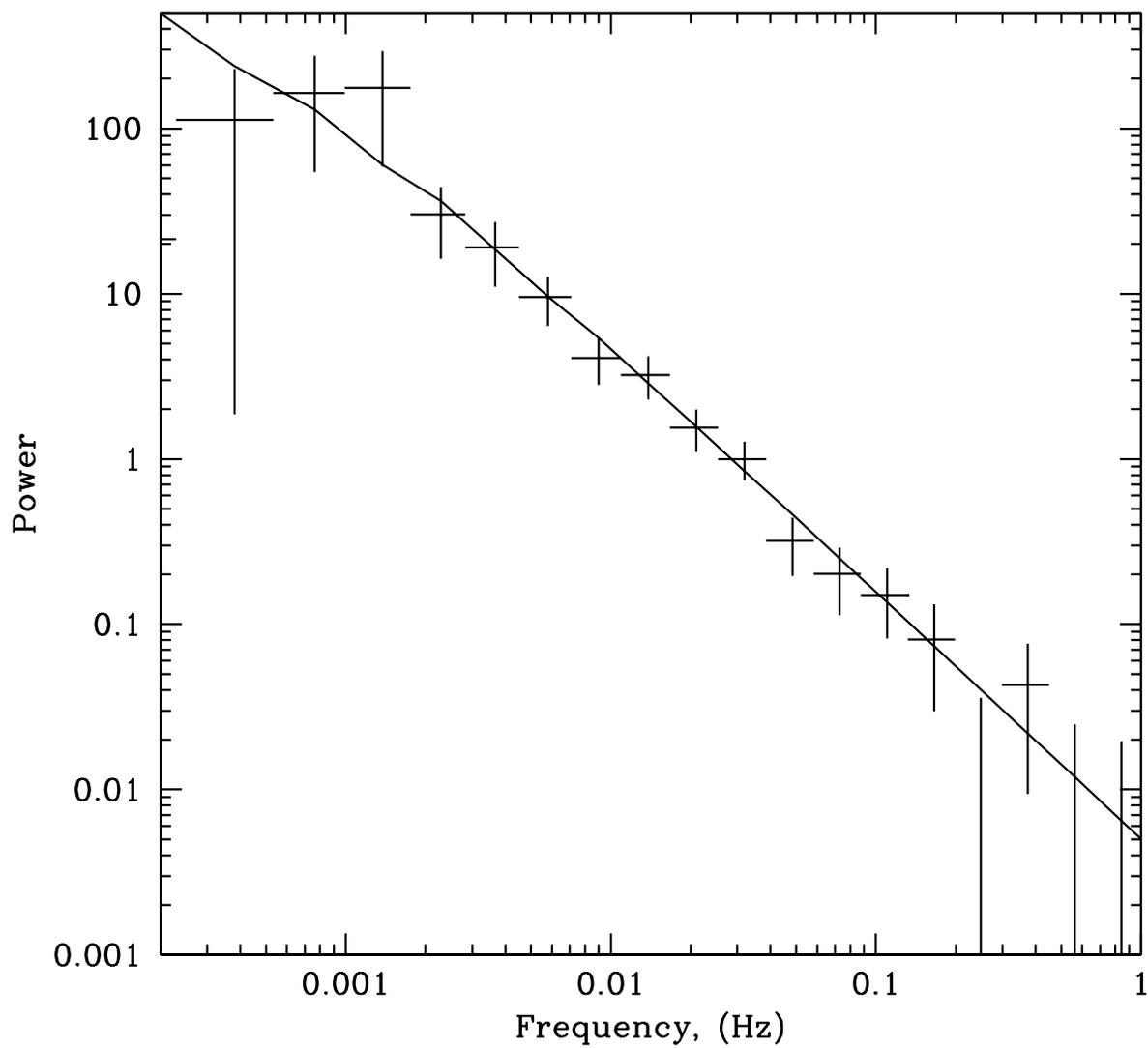}
\caption{Power spectrum of XTE J1901+014 obtained from PN/XMM-Newton data and fitted by a power law.}
\end{figure*}
\newpage

\begin{figure*}
\vbox{
\includegraphics[width=8.5cm]{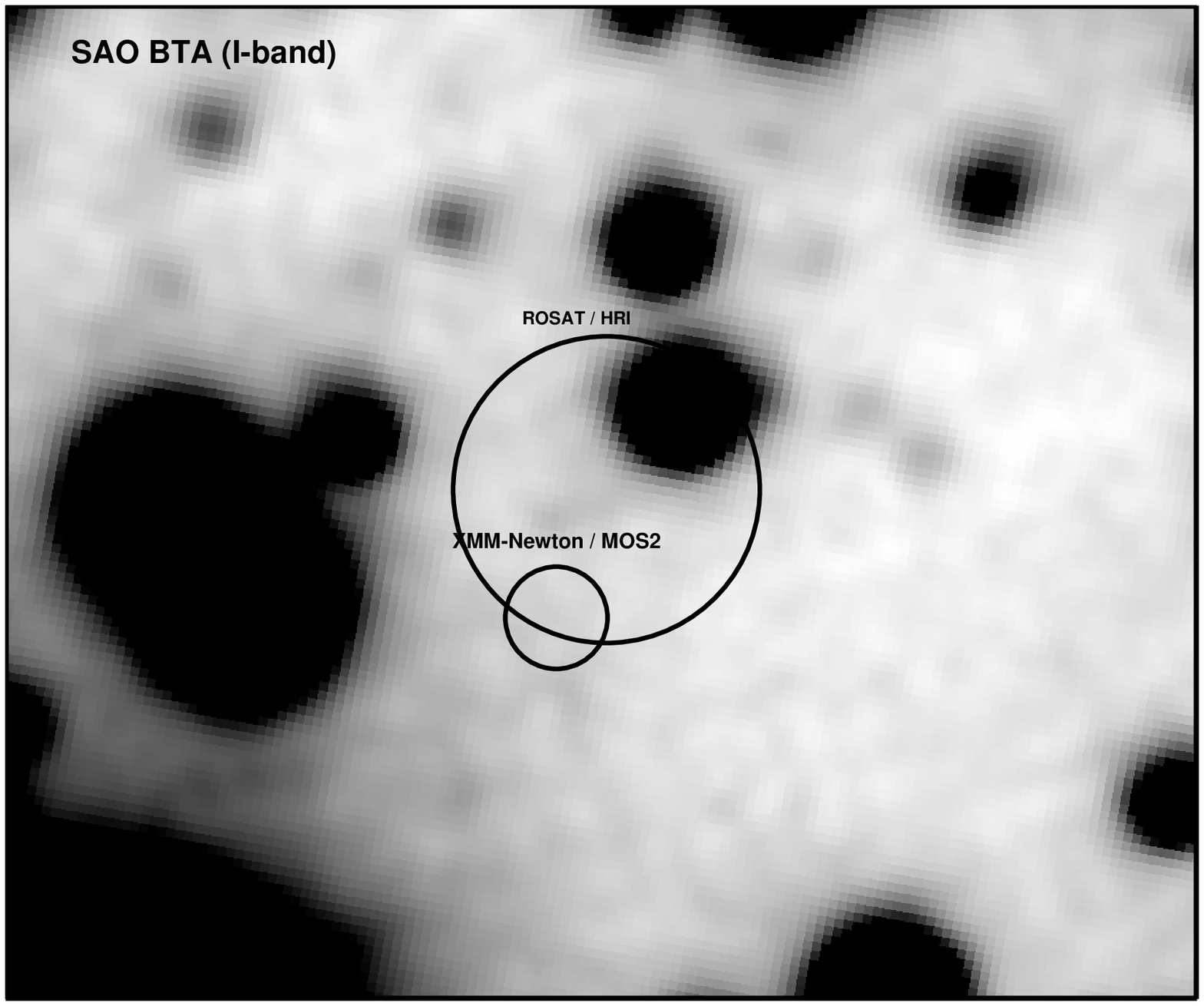}
\includegraphics[width=8.5cm]{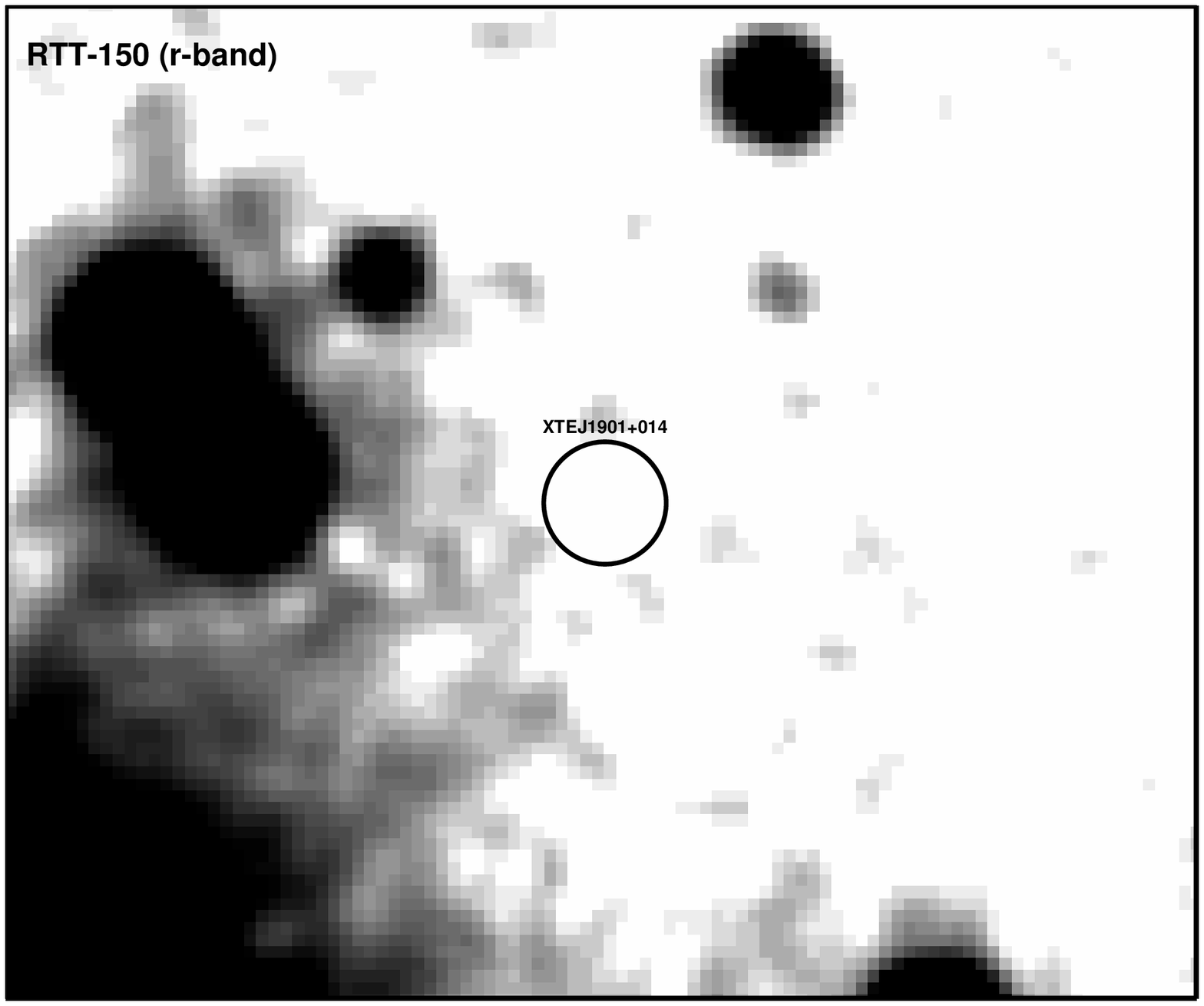}
\includegraphics[width=8.5cm]{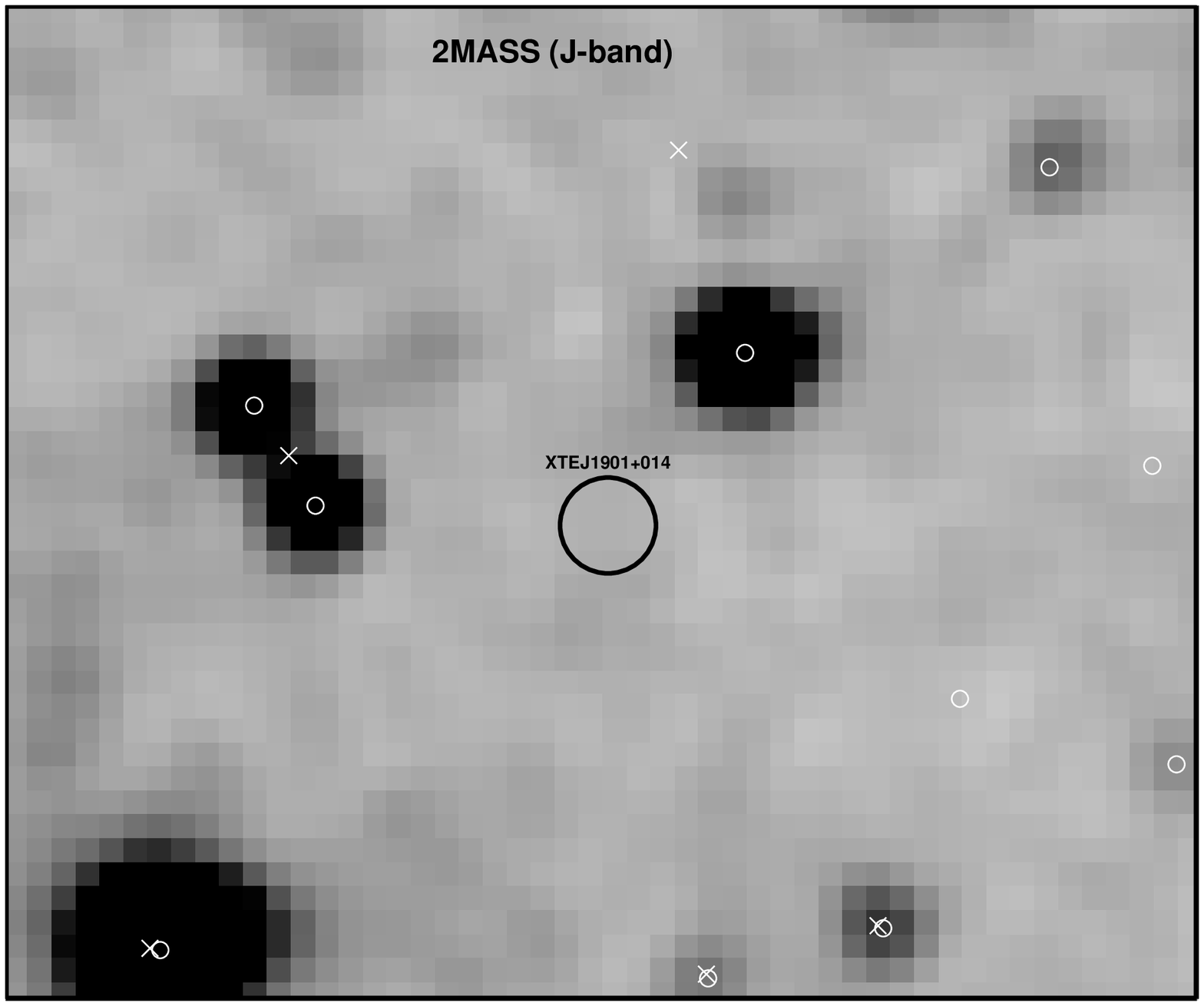}
}
\caption{Fig. 3. Images of the sky region containing XTE J1901+014: (a) in the I band based on BTA data, (b) in the r' band based on RTT-150 data; and (c) in the infrared J band based on 2MASS data. The localization center and the XMM-Newton/MOS2 (2\arcsec )and ROSAT/HRI (5\arcsec) error boxes of XTE J1901+014
are shown in the images. The circles and crosses mark
the positions of the nearest 2MASS and USNO-B1
sources.
}
\end{figure*}

\newpage

\begin{figure*}
\includegraphics[width=17cm]{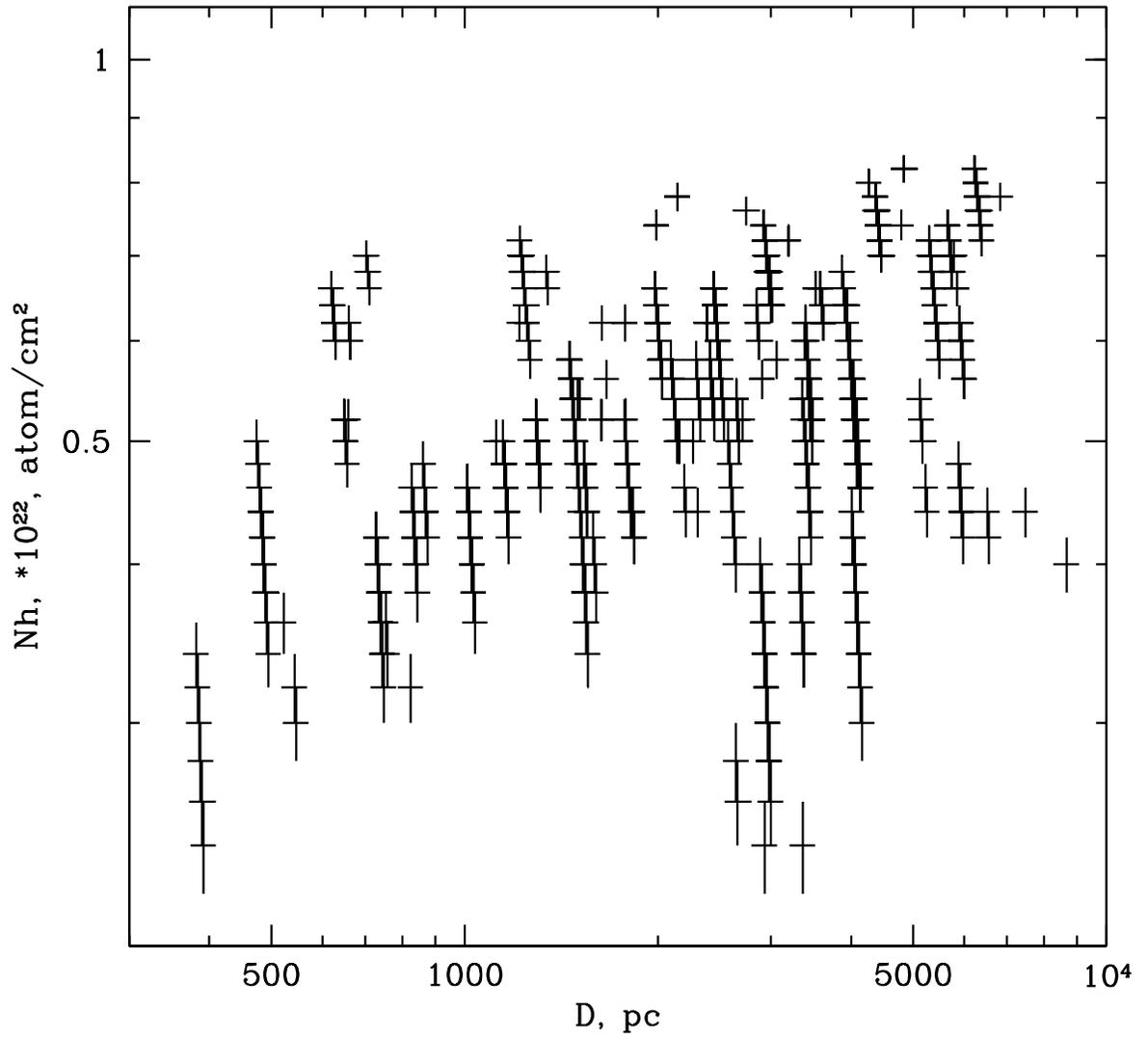}
\caption{Distance dependence of the absorption obtained from absorption and distance estimates for stars near XTE J1901+014 based on optical and infrared catalog data.}
\end{figure*}

\newpage

\begin{figure*}
\vbox{
\includegraphics[width=8.5cm]{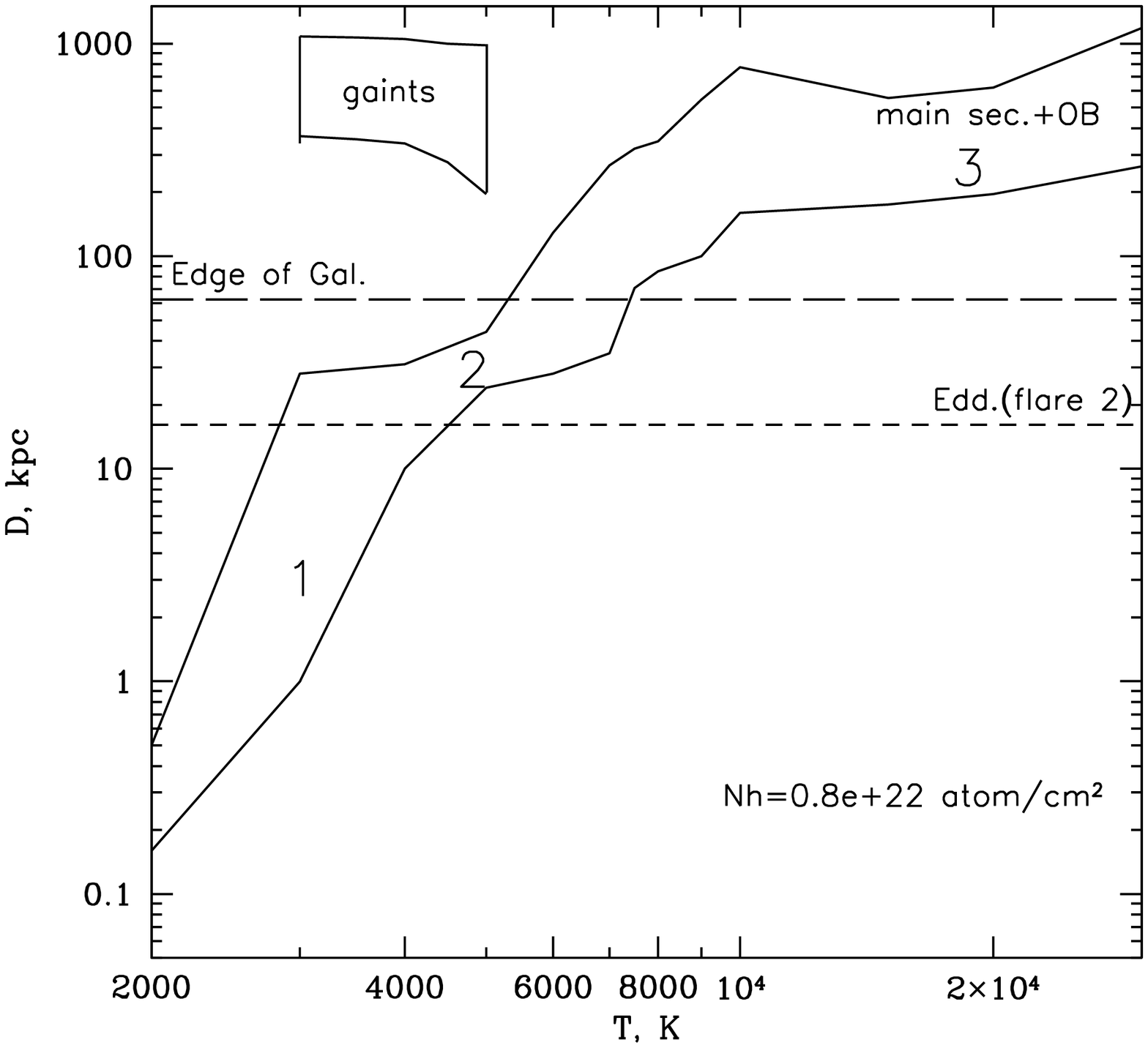}
\includegraphics[width=8.5cm]{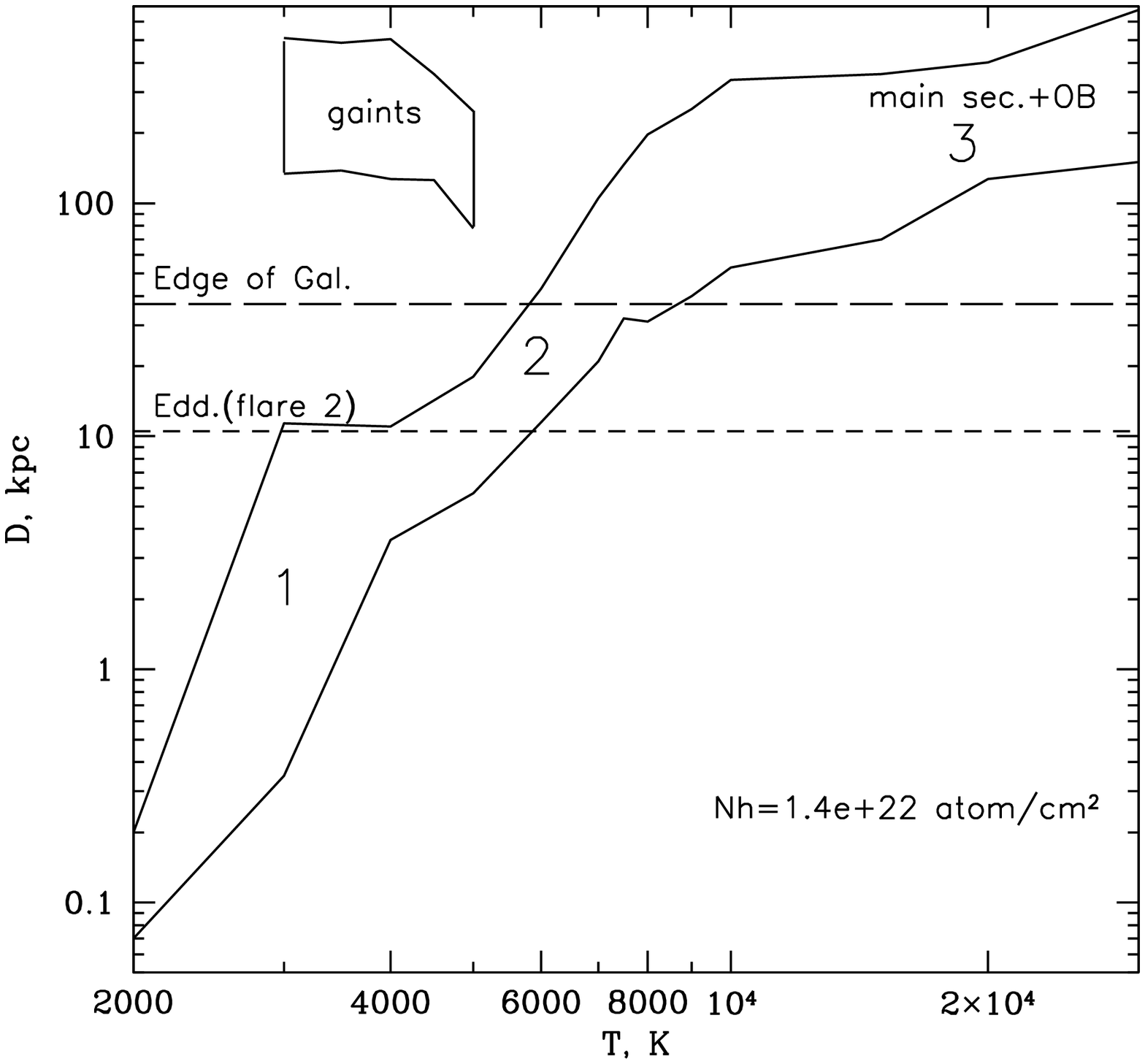}
\includegraphics[width=8.5cm]{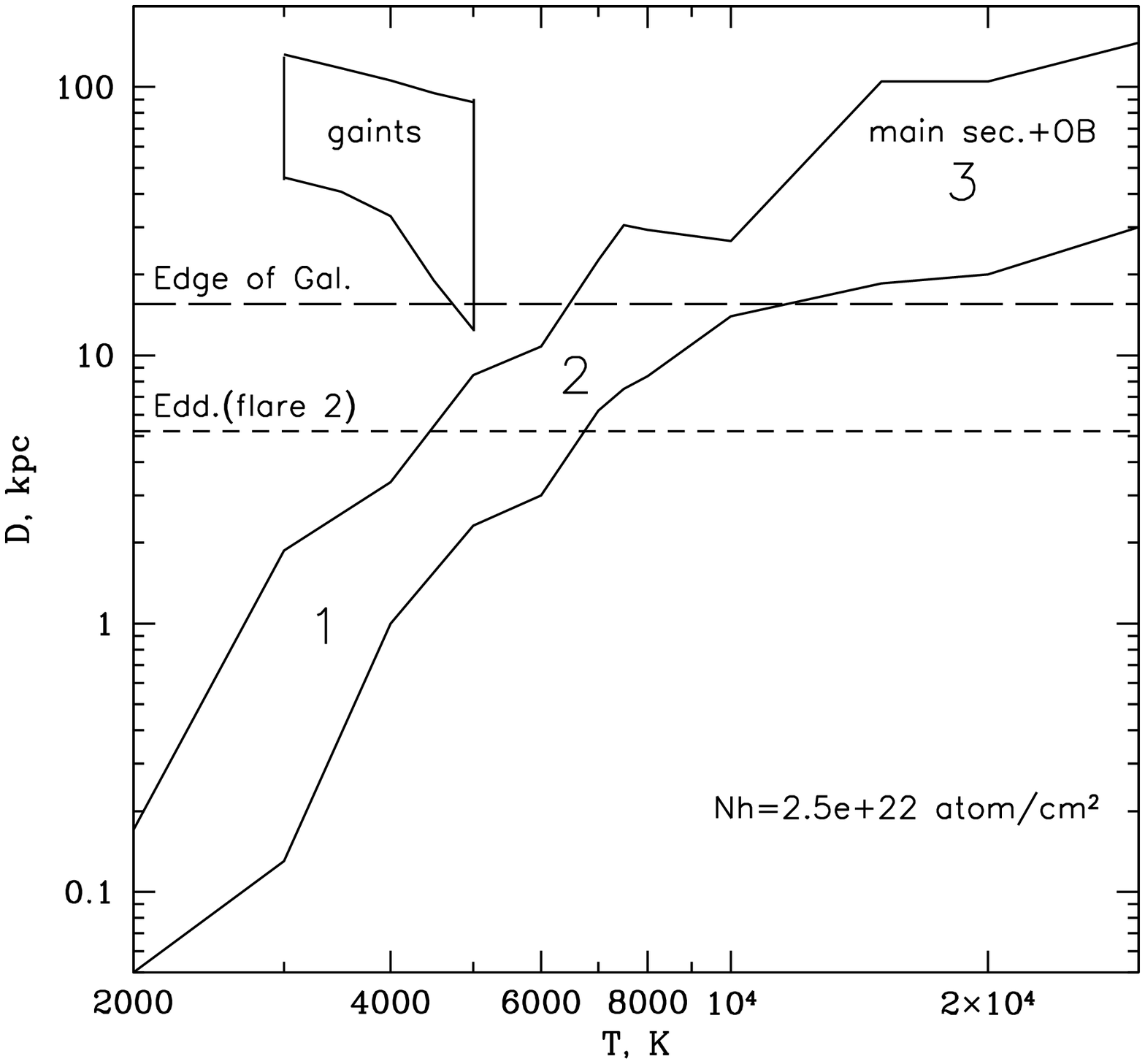}
}
\caption{Distances to the stars versus their temperature (type) for various values of the absorption. Main-sequence stars and
OB giants constitute the extended strip; the region of giants is also shown. The long dashed line marks the distance to the
Galactic edge in the direction under consideration; the short dashed line marks the distance at which the Eddington luminosity
limit for the neutron star during the XTE J1901+014 outburst in 2002 is reached. The numbers mark the following regions:
1, the star of this type can be the companion of XTE J1901+014, whose luminosity was below the Eddington limit during the
2002 outburst; 2, these stars can be the companions, but the luminosity during the outburst was above the Eddington limit; 3, these stars cannot be the companions.}
\end{figure*}

\newpage

\end{document}